\documentclass[aps,pra,preprint,
floatfix,
superscriptaddress,
onecolumn,
footinbib,
floatfix,
longbibliography]{revtex4-1}

\usepackage[T1]{fontenc}
\usepackage[utf8]{inputenc}

\usepackage{amsmath}
\usepackage{amssymb}

\usepackage[normalem]{ulem}

\usepackage[dvipsnames]{xcolor}

\usepackage{hyperref}
\hypersetup{colorlinks=true,citecolor=Cyan,urlcolor=Red}

\usepackage[squaren]{SIunits}

\usepackage[pdftex]{graphicx}
\usepackage{grffile} 

\usepackage{slashed}

\let\vec\boldsymbol

\newcommand{\ud}{\mathrm{d}}
\newcommand{\Ai}{\mathrm{Ai}}

\begin{document}

\title{Polarized QED cascades}

%
%

	\author{Daniel Seipt}
	\email{d.seipt@hi-jena.gsi.de}
	
	\affiliation{Helmholtz Institut Jena, Fr\"obelstieg 3, 07743 Jena, Germany}
	\affiliation{GSI Helmholtzzentrum f\"ur Schwerionenforschung GmbH, Planckstrasse 1, 64291 Darmstadt, Germany}
	\affiliation{The G\'erard Mourou Center for Ultrafast Optical Science, University of Michigan, Ann Arbor, Michigan 48109, USA}
	
	\author{Christopher P. Ridgers}
	\affiliation{York Plasma Institute, Department of Physics, University of York, York YO10 5DD, United Kingdom}

	\author{Dario Del Sorbo}
	\affiliation{High Energy Density Science Division, SLAC National Accelerator Laboratory, Menlo Park, CA 94025, USA}

	\author{Alec G. R. Thomas}
	\affiliation{The G\'erard Mourou Center for Ultrafast Optical Science, University of Michigan, Ann Arbor, Michigan 48109, USA}

    \begin{abstract}
    By taking the spin and polarization of the electrons, positrons and photons into account in the strong-field QED processes of nonlinear Compton emission and pair production, we find that the growth rate of QED cascades in 
    ultra-intense laser fields can be substantially reduced. While this means that fewer particles are produced, we also found them to be highly polarized. We further find that the high-energy tail of the particle spectra is polarized opposite to that expected from Sokolov-Ternov theory,
    which cannot be explained by just taking into account spin-asymmetries in the pair production process, but results significantly from ``spin-straggling''.
    We employ a kinetic equation approach for the electron, positron and photon distributions, each of them  spin/polarization-resolved, with the QED effects of photon emission and pair production modelled by a spin/polarization dependent Boltzmann-type collision operator. For photon-seeded cascades, depending on the photon polarization, we find an excess or a shortage of particle production in the early stages of cascade development, which provides a path towards a controlled experiment. Throughout this paper we focus on rotating electric field configuration, which represent an idealized model 
    and allows for a straightforward interpretation of the observed effects.
    \end{abstract}

    \date{\today}
    \maketitle

    \section{Introduction}
        
    The ongoing development of high-power petawatt class lasers \cite{Danson2019} has already opened new avenues for high-intensity laser-plasma physics. Already with present day laser technology the effect of QED process can be observed in high-intensity laser-matter interactions \cite{Cole:PRX2018,Poder:PRX2018,Blackburn:RevModPlasma2020}. With the upcoming generation of multi-10 PW laser systems \cite{Tanaka:MRE2020,SEL100PW} it is expected that laser-plasma interactions will enter a novel regime of QED plasma physics where strong-field QED process such as high-energy photon emission via nonlinear Compton scattering and electron-positron photo-production will play an important role for the overall dynamics of the plasmas \cite{Zhang:Perspective}.

    One of the most striking predictions of strong-field QED \cite{Ritus:JSLR1985,DiPiazza:RevModPhys2012} is the formation of \emph{avalanche}-type QED cascades \cite{Bell:PRL2008} at laser intensities approaching $10^{24}$ W/cm$^2$. In these cascades the prolific generation of high-energy photons and electron-positron pairs can convert an initially strong laser field into a hot and dense plasma of electrons, positrons and photons.  One of the conclusions drawn from these predictions was that the Schwinger field $E_S=\unit{1.3\times10^{18}}{\volt\per\metre}$ is then difficult to reach because of laser field depletion during the cascade \cite{Fedotov:PRL2010,Grismayer:PhysPlas2016,Seipt:PRL2017}.
    
    In a cascade, efficient pair production is facilitated by hard photons which have a quantum parameter $\chi_{\gamma} \gtrsim 1$. A particle with four-momentum $p^\mu$ has quantum parameter $\chi = e | F^{\mu\nu} p_\nu | / m^3 $, where $F^{\mu\nu}$ is the EM field strength tensor, $e$ the elementary charge and $m$ the electron mass. Those photons, in turn, are emitted by leptons, i.e. electrons and positrons, with $\chi_e \gtrsim 1$, leading to exponential increases in particle number \cite{Bell:PRL2008,Kirk:PPCF2009,Fedotov:PRL2010,Bulanov:PRL2010,Nerush:PRL2011,Nerush:PhysPlas2011,Elkina:PRSTAB2011,Slade-Lowther:NJP2019}.
    A renewable supply of sufficiently many leptons with $\chi_e \gtrsim 1$ necessarily can be provided if the electric field of the laser is capable of accelerating low-energy leptons with $\chi_e\ll 1$ to $\chi_e \gtrsim 1$. 
    
    This is one of the decisive features in \emph{avalanche-type} cascades, which can occur, for instance, in rotating electric fields at the magnetic nodes for two counter-propagating circularly polarized laser pulses \cite{Bell:PRL2008,Mironov2014}. 
    Such an avalanche-type cascade exhibits an exponential growth in particle number, limited  by the available (laser) field energy. By contrast, in \emph{shower}-type cascades that occur in interactions of of high-energy particle with non-accelerating field configurations, such as in collisions with (nearly) plane-wave laser pulses, the leptons are \emph{not} reaccelerated to $\chi_e \gtrsim 1$, and therefore the cascade multiplicity is limited by the maximum initial particle energy \cite{Bamber:PRD1999,Abramowicz2019,Meuren:2020}.

    QED pair-cascades are important in extreme astrophysical environments, where they may develop in matter, photon gas and magnetic field backgrounds \cite{Aharonian:AstroPart2003}. In the latter case of cascades in strong magnetic fields, as found in pulsar atmospheres  \cite{Sturrock:AstrophysJ1971,Ruderman:AstrophysJ1975}, the same basic theoretical framework to describe the quantum processes is used for describing laser-plasma interactions \cite{Ridgers:JCompPhys2014}.
    {There has been significant recent progress in understanding important physics details of cascades, including aspects of laser polarization, field structure \cite{Bashmakov:PhysPlas2014,Esirkepov:PLA2015,Grismayer:PRE2017} and seeding \cite{Jirka:PRE2016,Tamburini:2017}.}

    Lepton spin- and polarization effects in strong-field QED(-plasma) processes have gained some significant recent interest \cite{DelSorbo:PRA2017,DelSorbo:PPCF2018,Seipt:PRA2018,Seipt:PRA2019,Li:PRL2019,Chen:PRL2019,Wan:PLB2019,Thomas:2020}, (even though some fundamental scattering cross sections in strong laser fields were calculated much earlier \cite{Ivanov:EPJC2004,Ivanov:EPJC2005}). It was shown that electron beams colliding with bi-chromatic laser pulses can self-polarize due to hard-photon emission \cite{Seipt:PRA2019}, and that positrons generated by these photons are polarized as well \cite{Chen:PRL2019}. Alternatively, it was proposed to collide electron beams with elliptically polarized laser pulses to generate polarized lepton jets \cite{Li:PRL2019,Wan:PLB2019}. The radiative self-polarization of electrons in rotating electric fields in analogy to the Sokolov-Ternov effect \cite{book:Sokolov,Ternov:PhysUspekh1995} was studied e.g.~in \cite{DelSorbo:PRA2017,DelSorbo:PPCF2018}. An interesting scenario is the impact of particle polarization on the formation of avalanche-type cascades, which may be expected to potentially be a large effect due to its exponential nature.
    
    In this paper, we study the influence of lepton and photon spin and polarization on the formation of QED cascades. We find that the inclusion of particle polarization does affect the growth rates of cascades, and that all particle species in the developing cascades are highly polarized. Moreover, the high-energy tails of the lepton energy spectra are polarized opposite to what is expected from Sokolov-Ternov theory. In photon-seeded cascades we also find an excess of pairs in the early stages when the cascade is seeded by $\perp$-polarized photons (compared to unpolarized or $\parallel$-polarized photons), which presents a path towards an experimental confirmations of the theoretical results. 
    
    Our paper is organized as follows: In Section \ref{sect:model} we first develop the theoretical model based on a Boltzmann-type kinetic equation with a polarization dependent collision operator. In Section \ref{sect:results} numerical results are presented and discussed. Our findings are summarized and concluded in Section \ref{sect:conclusions}. Technical details and background are provided in the Appendix, including details on the numerical method.
    
    \section{Theoretical Model}
    \label{sect:model}
    
    So far, in simulations of cascades only the photon polarization was taken into account in \cite{King:PRA2013}, but lepton polarization was neglected. That means, in each generation the leptons emit polarized photons, but without any influence from previous generations because leptons are and stay unpolarized. Here, the full polarization evolution is taken into account consistently over many generations. The laser field is described as a rotating electric field $\vec E(t) = E_0 (\cos\omega t , \sin\omega t , 0)$, which is a suitable model for colliding laser pulses widely used in the literature \cite{Bell:PRL2008,Fedotov:PRL2010,Nerush:PhysPlas2011,Elkina:PRSTAB2011,Blinne2014,Mironov2014,Kohlfurst:PRD2019} and is generated at the magnetic nodes of colliding laser pulses. The rotating electric field represents a simplified model of a real high-intensity laser plasma interaction which allows for a straightforward analysis of the observed particle polarization properties.

    We model evolution of the polarized QED cascades by a Boltzmann-type kinetic equation \cite{Nerush:PhysPlas2011,Elkina:PRSTAB2011,Neitz:PRL2013,Bulanov:PRA2013} for the one-particle distribution functions $f_q^s$ of of electrons ($q=-1$), positrons ($q=+1$) and photons ($q=0$), where $q$ is the particle's charge in units of $e$, and in a polarization state $s$. 
    {The transport equations for $f_q^s$ in a rotating electric background field are derived in \ref{app:A}.}
    At high intensity, $a_0\gg1$, {where $a_0=eE_0/m\omega$ is the normalized laser vector potential,} the coherence length for the quantum processes $\lambda/a_0$ becomes much shorter than the laser wavelength $\lambda$ \cite{Ritus:JSLR1985}. This scale separation allows for the quantum processes to be described as local collisions by a Boltzmann collision operator, where the rates for quantum processes calculated in a constant crossed field. In the rotating electric field, the leptons can be spin-polarized up ($\uparrow$) or down ($\downarrow$) along the spin-quantization axis $\vec e_z \parallel \vec E\times \vec p$, which is the magnetic field direction in the rest frame of the leptons. This represents a global non-precessing spin quantization axis which means that the spin vectors of particles polarized along the z-axis do not precess according to the T-BMT equation \cite{Bargmann:PRL1959,DelSorbo:PRA2017}. (The spin-polarization components perpendicular to $\vec e_z$ do not have to be treated explicitly, see \ref{app:A} for details.)
    In our model the background field is homogeneous so Stern-Gerlach forces are {not present}. Further, they can be neglected because they are much weaker than the Lorentz force {if field gradients occur on length scales much longer than the Compton wavelength} \cite{Ekman:PRE2017,DelSorbo:PPCF2018,Thomas:2020}. Quantum radiation reaction and its spin dependence is included in the calculations automatically through the photon emission. Photons can be in a polarization eigenstate $\parallel$ or $\perp$ to the plane of the rotating electric field. Additional details can be found in \ref{app:A}.

    Throughout the paper we use natural Heaviside-Lorentz units with $\hbar=c=\epsilon_0=1$. We further introduce normalized (dimensionless) time $\omega t \to t$, momentum $\vec p/m \to \vec p$
    and electric field $e \vec E / m\omega  \to \vec E$.
    {Thus, the normalized $E_0$ has the same numerical value as $a_0$. For the Minkowski metric we use the sign convention $(+,-,-,-)$.}

    The normalized Boltzmann equation can then be written as 
    \begin{align}  \label{eq:Boltzmann-normalized}
        \left( 
            \frac{\partial}{\partial t} + q  \vec E \cdot \nabla_{\vec p}  
        \right) f_{q}^{s}(\vec p,t) 
        = 
        \mathcal C_q^s[\{ f_{q'}^{s'} \}] \,,
    \end{align}
    where $\mathcal C^s_q$ are the collision operators describing all relevant strong-field QED processes, with the charge $q$ labelling different particle species. The  $\mathcal C^s_q$ are linear functionals of the spin and polarization dependent differential rates for nonlinear Compton scattering and pair production \cite{Nikishov:JETP1964b,Ritus:JSLR1985,Seipt:PRA2011,King:PRA2013}.
    The rates were calculated from first-order high-intensity QED Feynman diagrams in the Furry picture within the local constant crossed field approximation  \cite{Ilderton:PRA2019,Blackburn:PRA2020,Seipt-King}{, see also \ref{app:rates}.}

    The lepton collision operators $\mathcal C^{s}_{\pm 1}$ describe the energy losses due to radiation emission and accompanying spin-flip transitions, as well as a gain term due to pair production by photons. The photon collision operator $\mathcal C_0^j$ contains terms for the absorption of $j$-polarized photons during pair production and the generation of photons by non-linear Compton emission off of electrons and positrons. The explicit expressions for the polarization dependent collision operators are given below in Section \ref{sect:CO}.

    The classical advection operator on the left hand side of Eq.~\eqref{eq:Boltzmann-normalized} does not mix different polarization states (because of the choice of the non-precessing spin quantization axis, see \ref{app:A}). Only the Boltzmann collision operator on the r.h.s. of Eq.~\eqref{eq:Boltzmann-normalized} mixes different spin states and, thus, can lead to a change of the polarization of the particles. There is no mixing of different photon polarization states due to vacuum polarization effects since the $\parallel/\perp$ states are eigenstates of the photon polarization tensor \cite{Karbstein:PRD2013,King:PRA2016}.

   	\subsection{Characteristics}
   
   The characteristics of the {left-hand-side} of \eqref{eq:Boltzmann-normalized} describing the classical {trajectory} can be found analytically by solving {the equation of motion in the rotating electric field:}
   \begin{align} \label{eq:orbit}
   \vec p(t) = q \int_{t_0}^t \ud t' \, \vec E(t')  + \vec p(t_0) \,,
   \end{align}
   which means{, for leptons with $q=\pm1$,}  $\vec p(t) = (p_x(t),p_y(t),0)$ lies on a circle with radius $a_0$ around the centre $\vec P = \vec p (t_0) + q \vec a(t_0)$, where
   $\vec a = - \int \! \ud t \: \vec E$ is the normalized vector potential and 
   $\vec P$ is the conserved canonical momentum. Leptons starting with zero momentum can be accelerated up to {$|\vec p|=2a_0$}, where $a_0 = e E_0 /m\omega$. 
   For the photons, with $q=0$, Eq.~\eqref{eq:orbit} is just free ballistic motion with constant momentum.

   For a numerical solution of the Boltzmann equation it is convenient to 
   {introduce the angle $\varphi$ between $\vec E$ and $\vec p$ for all particle species \cite{Nerush:PhysPlas2011}.}
   The equations of motion for the characteristics {in terms of the new variables $p=|\vec p|$ and $\varphi$} read
   \begin{align}
   \frac{\ud p}{\ud t} & = q a_0 \cos \varphi \,, \\
   \frac{\ud \varphi}{\ud t} & = - q \frac{a_0}{p} \sin\varphi - 1 \,,
   \end{align}
   with the completely analytical solution,
   \begin{align}
   p &= \sqrt{  p_0^2 
   	+ c_0^2 
   	+ 2 q c_0 p_0  \cos\left( \varphi_0 - \frac{\Delta t}{2}\right)
   }
   \label{eq:characteristics-S1} \,,\\
   \varphi &=
   \arctan \left(
   \frac{p_0 \sin \varphi_0 + q c_0 \sin \frac{\Delta t}{2}}
   {p_0 \cos \varphi_0 + q c_0 \cos \frac{\Delta t}{2}}  \right) - \Delta t \,,
   \label{eq:characteristics-S2}
   \end{align}
   with $c_0 = 2 a_0 \sin \frac{\Delta t}{2}$. Here $p_0,\varphi_0$ means those quantities are taken at $t_0$. The characteristics are written in terms of the time-step $\Delta t = t- t_0$. Note that Eqs.~\eqref{eq:characteristics-S1} and \eqref{eq:characteristics-S2} are exact to all orders in $\Delta t$. For the sake of completeness, we also write down the Boltzmann equation in the {new variables}:
   \begin{align}
   \left [ \frac{\partial }{\partial t} + q a_0 \cos \varphi \frac{\partial}{\partial p} - \left( q\frac{a_0}{p} \sin \varphi +1\right) \frac{\partial }{\partial \varphi } \right ] f_{q}^{s} = \mathcal C_q^s \,.
   \end{align}

    \subsection{Polarization dependent collision operators}
    \label{sect:CO}
    	
    The benefit of {introducing $\varphi$} becomes apparent when considering the quantum transitions; ultrarelativistic particles emitted during the quantum processes are kinematically restricted to be collinear to the emitting particle and therefore involve transitions at constant $\varphi$ such that only the magnitude of momentum, $p$, changes. The angle $\varphi$ enters the collision operator only parametrically via the quantum parameters
    \begin{align} \label{eq:chi}
        {\chi_q = \frac{a_0}{a_S}  \sqrt{ q^2 + p^2 \sin^2\varphi}} \,,
    \end{align}
    which determine the probabilities of the QED processes, and with 
    the normalized Schwinger vector potential $a_S = m/\omega$.

    {Quantum} transitions leave the {momentum} angle $\varphi$ unchanged. This greatly simplifies the numeric calculations. The polarization dependent collision operators for electrons, positrons and photons can be written down as
    \begin{align} 
    	\label{eq:collision-operator-e}
    	\mathcal C_{-1}^{s}(p)
    	& =	- f_{-1}^{s}(p) \sum_{s',j} W_{-1}^{ss'j}(p)
    	+ \sum_{j,s'} \int_p^\infty \!\ud p' \: \frac{p'}{p} \: f_{- 1}^{s'}(p') \, w^{s's j}_{-1}(p'\to p) 
    	\nonumber \\
    	& \qquad \qquad \qquad 
    	+ \sum_{j,s'} \int_p^\infty \!\ud k \: \frac{k}{p}\: f_{0}^{j} (k) \, w^{s's j}_{0}(k\to k- p)
    	\,,\\
    	\label{eq:collision-operator-p}
    	\mathcal C_{+ 1}^{s}(p)
    	& =	- f_{+1}^{s}(p) \sum_{s',j} W_{+1}^{ s  s' j}(p)
    	+ \sum_{j,s'} \int_p^\infty \!\ud p' \: \frac{p'}{p} \: f_{+ 1}^{s'}(p') \, w^{ s' sj}_{+1}(p'\to p) 
    	\nonumber \\
        & \qquad \qquad \qquad 
    	+ \sum_{j,s'} \int_p^\infty \!\ud k \: \frac{k}{p}\: f_{0}^{j} (k) \, w^{ss'j}_{0}(k\to p)
    	\,,\\%
    	\label{eq:collision-operator-g}
    	\mathcal C_{0}^{j}(k)
    	& = - f_{0}^{j}(k) \sum_{s,s'} W^{ss' j}_0 (k)
    	+ \sum_{q=\pm 1} \sum_{s,s'} \int_k^\infty \!\ud p \: \frac{p}{k} f_{q}^{s}(p)  \, w^{ss'j}_{q}(p\to p-k)
    \end{align}
    The collision operators \eqref{eq:collision-operator-e}--\eqref{eq:collision-operator-g}    are a generalization of both the collision operators to describe the evolution of unpolarized cascades \cite{Nerush:PhysPlas2011,Elkina:PRSTAB2011,Bulanov:PRA2013,Neitz:PRL2013}, as well as the collision operator for the radiative spin-polarization of electron beams \cite{Seipt:PRA2019}. They are functionals of the differential rates for nonlinear Compton scattering $w_{\pm1}$ and nonlinear Breit-Wheeler pair production by a photon $w_0$, as well as the corresponding total (momentum integrated) rates (see \ref{app:rates} for details)
    \begin{align}
     W_{q}^{ss'j}(p) = \int_{0}^p \! \ud k  \: w_{q}^{ss'j}(p\to k) \,,    
    \end{align}
    as well as the polarized one-particle distribution functions of all particle species $f_q^s$.  We should emphasize again that these collision operators furnish transitions between different momentum of the particles, but only with regard to their magnitude. The angle $\varphi$ enters only parametrically, and the quantum rates depend on $\varphi$ only via their respective quantum parameters $\chi_q(p,\varphi)$, Eq.~\eqref{eq:chi}. {In that sense, these collision operators are effectively 1D (i.e. one momentum integral) because the momentum direction does not change in a quantum transition in the ultrarelativistic approximation (when collinear emission is assumed as usual).
    The momentum ratios, such as $p/k$ appearing in the last term of Eq.~\eqref{eq:collision-operator-g}, are reflections of this reduced dimensionality.
    The details of the reduction of the full 3D collision operator to the 1D ones for the rotating field case are discussed in the literature, see e.g.~Ref.~\cite{Elkina:PRSTAB2011,Nerush:PhysPlas2011}.
    }
    Let us now briefly discuss the physical meaning of the individual terms in the collision operators.
    	
    The first two terms in each of Eqns.~\eqref{eq:collision-operator-e} and \eqref{eq:collision-operator-p} describe the radiative energy loss of the leptons during photon emission, i.e.~quantum radiation reaction \cite{Elkina:PRSTAB2011,Niel:PRE2017,Ridgers:JPP2017}. These terms include both the possibility of spin-flip and non-flip transitions and thus are also responsible for the radiative lepton polarization, and they had been used already in \cite{Seipt:PRA2019}. 
    In addition, {they describe} the spin-dependence of quantum radiation reaction. The last term in each of \eqref{eq:collision-operator-e} and \eqref{eq:collision-operator-p} is for the generation of up/down polarized electrons and positrons from photons of arbitrary polarization, respectively.
    	
    For the photon collision operator $\mathcal C_0^j(k)$ in Eq.~\eqref{eq:collision-operator-g} the first term describes the absorption of $j$-polarized photons during pair production and the second term {describes} the generation of $j$-polarized photons by non-linear Compton emission {from} both electrons and positrons (sum{med} over charge $q$).

    \section{Results and Discussion}
    \label{sect:results}    
    
    \begin{figure}
    \begin{center}
	\includegraphics[width=0.8\columnwidth]{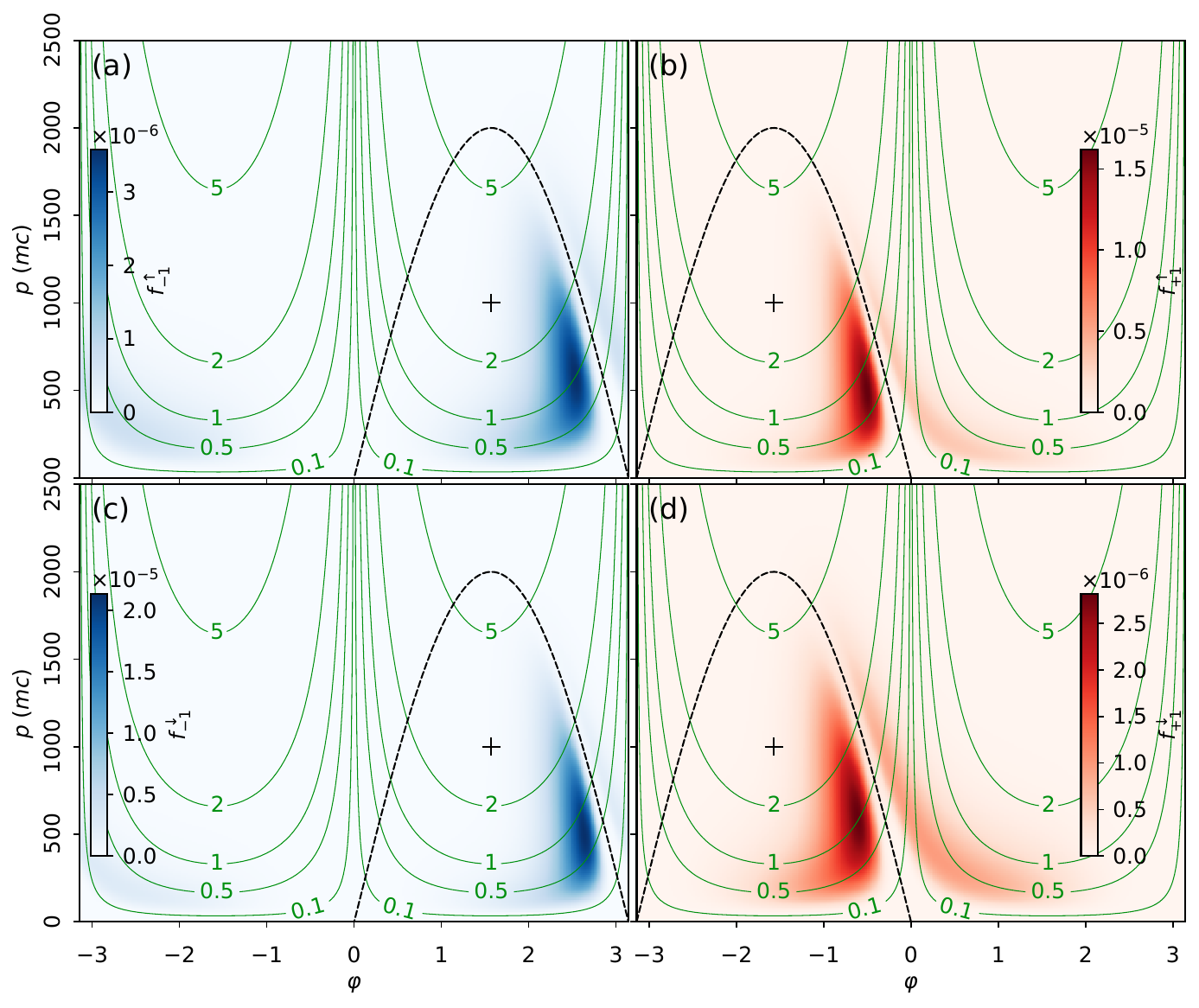}
    \end{center}
    \caption{Snapshot of the electron (a,c) and positron (b,c) distribution functions in an up (a,b) or down (c,d) polarization state for $a_0 = 10^3$ and $\omega t= 10$ in a rotating radial frame. Green curves are $\chi$ isocontours. Black dashed curves represent the separatrix of the classical advection $p = - 2 q a_0  \sin \varphi $, and black crosses are the corresponding fixed points at {$\varphi= -q \pi/2$, $p=a_0$}.}
    \label{fig:snapshot}
    \end{figure}
   
    Numeric solutions of Eq.~\eqref{eq:Boltzmann-normalized} for the up and down electron and positron distributions in the presence of both the quantum processes and the classical motion are shown in Fig.~\ref{fig:snapshot}, for $a_0 = 1000$ and $\omega = 1.55$~eV. In Fig.~\ref{fig:snapshot} we also plot the $\chi_{\pm1}$ isocontours, which greatly vary over the phase space. In particular, at the lines $\varphi = \pm\pi/2$, where the particle momentum and the electric field are perpendicular, the quantum parameters are $\chi_q\approx a_0 p/a_S$. A characteristic value of $\chi_q$ can be defined at the fixed points, {$\varphi= -q \pi/2$, $p=a_0$}, of the lepton phase-spaces as $\chi_\mathrm{FP} \simeq a_0^2/a_S$. For the parameters used in Fig.~\ref{fig:snapshot}, $\chi_\mathrm{FP}=3.03$.
    
    Photons are predominantly generated with $\parallel$-polarization. Leptons of both polarization states are dominantly produced by photons that reach $\varphi = \pm \pi/2$ because $\chi_0$ is largest there. Leptons produced at $\varphi = q \pi/2$ (i.e.~outside the separatrix) are accelerated to large $p$ by the electric field as they orbit to $\varphi = - q\pi/2$ and hence large values of $\chi_{\pm1}$ exceeding $2a_0^2/a_S$. When $\chi_{\pm1}\gtrsim 1$, photon emission is efficient, and therefore leptons, while they are being accelerated, are most likely to radiate photons. As a consequence, they lose sufficient energy to enter the closed orbits inside the separatrix, where they accumulate. Note that no classical orbits cross the separatrix and the quantum transitions only take leptons from outside the separatrix to inside, so once inside the separatrix, leptons are trapped. Spin-flip transitions occurring during photon emission lead to further polarization of the leptons. A movie of the evolution of the distribution functions is provided in the supplementary material.

    The shape of the distribution and their location inside the separatrix are determined by a balance of classical acceleration, feeding due to pair production, and photon emission. The latter two processes are spin dependent. Therefore, it is not surprising that the shapes of the up and down particle distributions are different. For instance, the values of $\chi_{\pm1}$ at the maximum of the distributions are $1.04$ for up electrons, but only $0.70$ for down electrons. The position of the distribution's peak, $\hat \varphi$, is also an indicator for the magnitude of radiation reaction effects. In Fig.~\ref{fig:snapshot} we find a difference in the angular shift of $ \hat \varphi_\downarrow - \hat \varphi_\uparrow \simeq 0.1$, signalling a spin-dependence of radiation reaction. By treating radiation reaction semiclassically as a continuous friction force (and neglecting pair production) \cite{Bulanov:PRE2011,Zhang:NJP2015}, the angle deviation is proportional to the strength of the radiation reaction force in a lowest order perturbative analysis.
    
    \begin{figure}
    \begin{center}
	\includegraphics[width=\columnwidth]{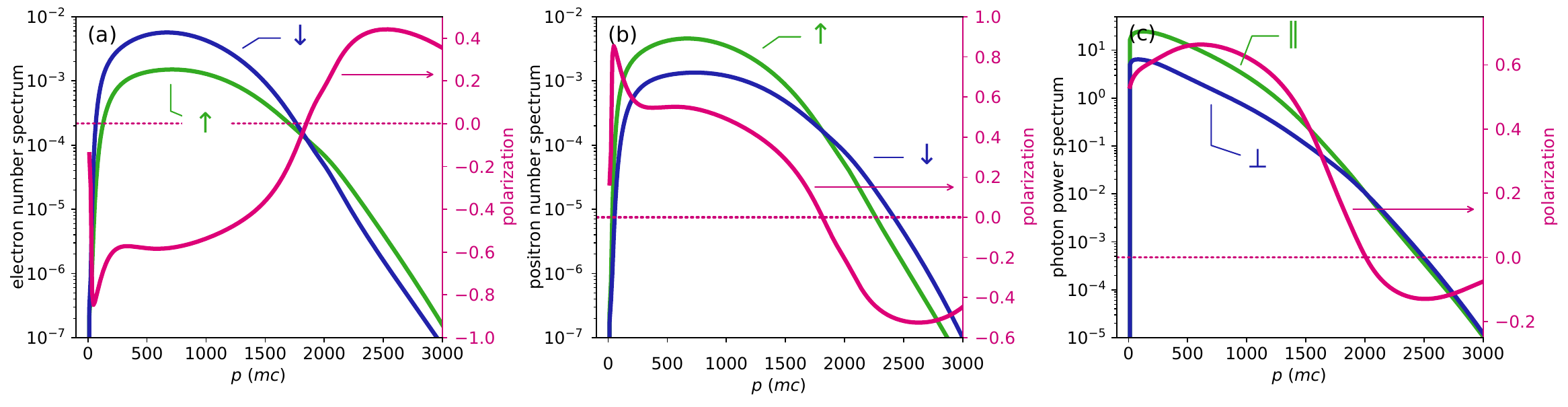}
    \end{center}
    \caption{{Momentum spectra (blue / green curves) and differential degree of polarization (pink curve) for electrons (a), positrons (b) and photons (c).} Same parameters as in Fig.~\ref{fig:snapshot}.}
    \label{fig:spectra}
    \end{figure}
    
    Figure~\ref{fig:spectra} shows the spectra of electrons (a), positrons (b) and photons (c) at $\omega t = 10$ for $a_0=1000$, seeded by an unpolarized electron. The plots also show the degree of polarization for each case, i.e.~$\int (f_{\pm1}^\uparrow - f_{\pm1}^\downarrow) \ud \varphi/\int (f^\uparrow_{\pm1}+f^\downarrow_{\pm1}) \ud \varphi$ for electrons and
    $\int (f_{0}^\parallel - f_{0}^\perp) \ud \varphi/\int (f^\parallel_{0}+f^\perp_{0}) \ud \varphi$ for photons. These {plots show} that the positron polarization is opposite to the electron polarization. Slight deviations occur because the cascade is seeded by electrons.
    For the leptons the main peak extends roughly up to about $2a_0$, which mostly corresponds to particles accumulating inside the separatrix. Those particles are dominantly polarized, as predicted by Sokolov-Ternov theory, with electrons more likely in a spin-down state (and positrons in an up-state). 
    {Interestingly, in the high-energy tail above $p\gtrsim 2a_0$ the electrons and positrons are \emph{oppositely polarized to the expected Sokolov-Ternov polarization}. 
    Looking at the spin and polarization dependent pair production spectra one finds that there is indeed an asymmetry in the lepton spectra where parallel polarized photons tend to produce up electrons (down positrons) at larger energies than the opposite polarization state \cite{Seipt-King}. However, this is only a small contribution to the observed effect in Figure \ref{fig:spectra}.
    The majority of the effect can be explained by the phenomenon of ``spin-straggling'' where one polarization state is more likely to reach the highest values of $\chi$, analogous to straggling effects previously observed without spin \cite{Blackburn:PRL2014}.
    Straggling means that the highest energy leptons in the spectrum most likely have not emitted a photon since their creation by the decay of a previously generated photon. For leptons with $\chi\sim 1$ the emitted photons carry a large fraction of the lepton's momentum such that at each emission the leptons lose a large fraction of their momentum.
    The photon emission rate for up electrons, $\sum_{s'j} W_{-1}^{\uparrow s'j}$, is smaller than that for down electrons, $\sum_{s'j} W_{-1}^{\downarrow s'j}$, such that they are less likely to emit and hence have a larger probability to reach higher energies than down electrons.
    This is contrary to the explanation of the Sokolov-Ternov effect, where the difference of spin-flip transition rates $\sum_{j} W_{-1}^{\downarrow \uparrow j}$ and $\sum_{j} W_{-1}^{\uparrow \downarrow  j}$ was decisive \cite{book:Sokolov}.
    }

    \begin{figure}
    \centering
    \includegraphics[width=0.9\columnwidth]{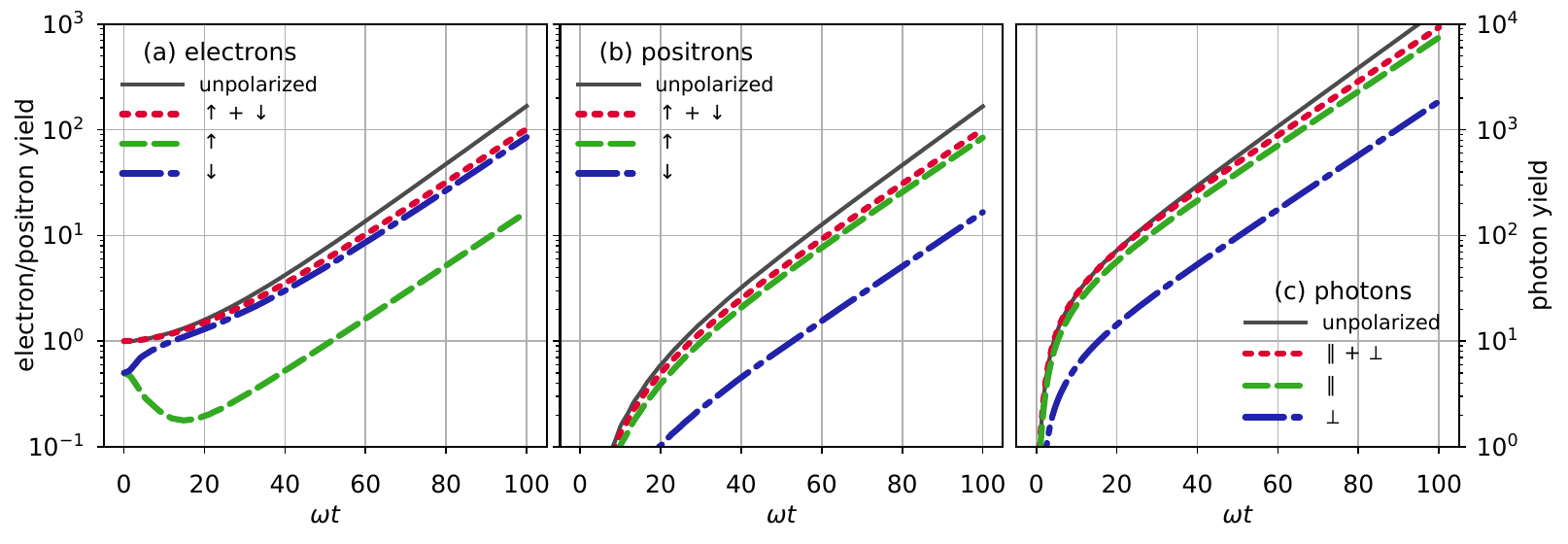}
    \caption{Time evolution of electron (a), positron (b)
    and photon (c) yields during a cascade seeded by unpolarized electrons, and for  $a_0=600$ and $\omega = \unit{1.55}{\electronvolt}$. Results for a polarized cascade are compared to an unpolarized cascade simulation.
    }
    \label{fig:yields}
    \end{figure}

    The important findings of previous studies of cascades \cite{Bell:PRL2008,Fedotov:PRL2010,Elkina:PRSTAB2011,Nerush:PhysPlas2011} was that they eventually reach exponential growth in particle number, $n = \int \! \ud \varphi \ud p \, p \: f \propto e^{\Gamma t}$. 
    {Our numerically calculated growth rates are in reasonable agreement with the analytical models of \cite{Grismayer:PRE2017,Kostyukov:2018}.}
    The evolution of the particle yields are shown in Fig.~\ref{fig:yields} for an unpolarized electron seeded cascade for parameters $a_0=600$ and $\omega = \unit{1.55}{\electronvolt}$. We compare a calculation of a polarized cascade with an unpolarized cascade, simulated using only unpolarized distribution functions and rates in the collision operator (see e.g.~\cite{Nerush:PhysPlas2011}). Fig.~\ref{fig:yields} (a) shows that the yields of up electrons decreases initially. This is because of spin flip transitions in the photon emission (Sokolov-Ternov effect). Later in the evolution of the cascade, pair production becomes more relevant and all particle yields enter an exponential growth phase, eventually reaching a common growth rate $\Gamma$.
    In the exponential growth phase there are $5$ times more down than up electrons (opposite for positrons). Moreover, a factor of $4$ more $\parallel$-polarized photons are emitted compared to $\perp$-polarized photons. {Thus, the particle distributions produced in a QED cascade are highly polarized.}

    \begin{figure}[!ht]
    \centering
    \includegraphics[width=\columnwidth]{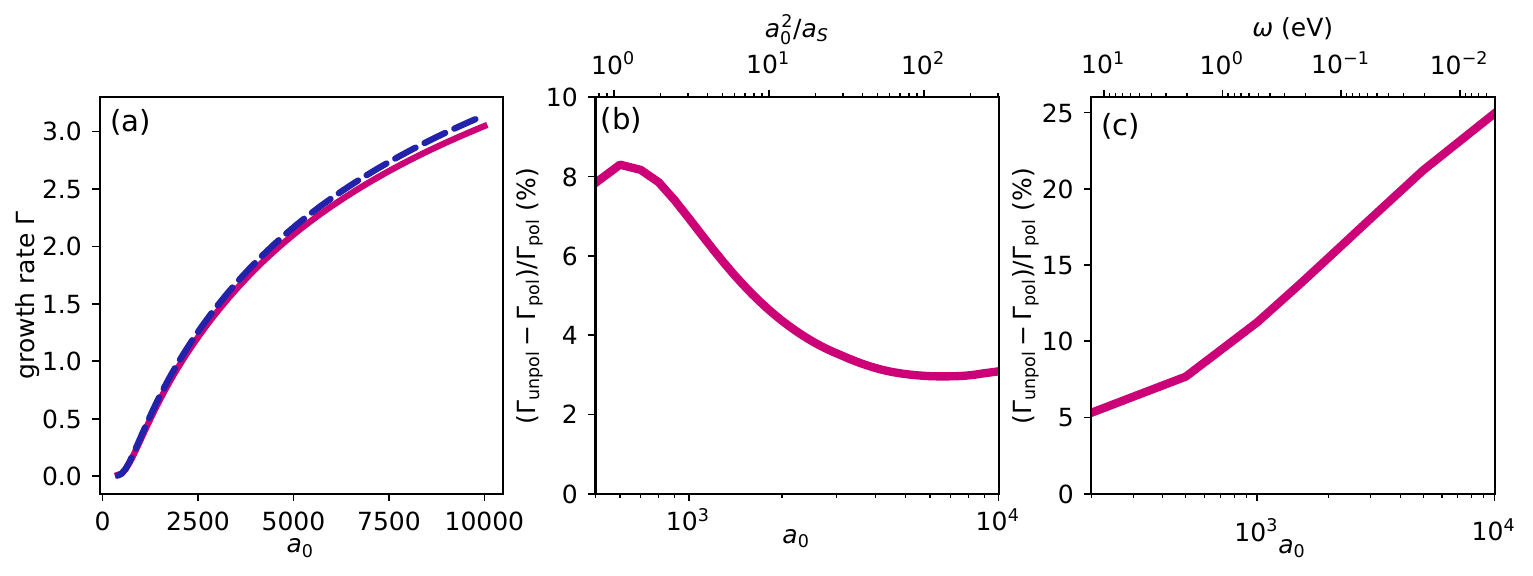}
    \caption{Cascade growth-rates $\Gamma$ (a) as a function of $a_0$ at fixed $\omega=\unit{1.55}{\electronvolt}$ and the {corresponding
    relative difference between polarized and unpolarized growth-rates
    (b), and for $\chi_\mathrm{FP}  =  1 = \mathrm{const.}$ (c).}
    }
    \label{fig:rates}
    \end{figure}

    The growth rate is calculated as $\Gamma_{q}^s = \ud \ln n_{qs} / \ud t$, with $\Gamma_{q}^s \to \Gamma$ as $t\to \infty$, and shown as a function of $a_0$ in Fig.~\ref{fig:rates} (a). We find that the growth rate of a polarized cascade when electron spin and photon polarization are properly taken into account is typically smaller than in an unpolarized cascade calculation by about $3\%$ to $8\%$, with the maximum discrepancy around $a_0=600$ in this case, {see Fig.~\ref{fig:rates} (b)}. It is worth noting that a reduction in the growth rate of $\sim$5\% at a growth rate of $\sim1$ (i.e.~$a_0\sim2000$) corresponds to a reduction in particle yield by an order of magnitude in less than 10 (laser) cycles. For fixed laser frequency $\omega$, the effective value of $\chi$ at the fixed point of the classical advection $\chi_\mathrm{FP} = a_0^2 \Omega$. It is also of interest to examine the behaviour at fixed 
    {$\chi_\mathrm{FP}=a_0^2 /a_S =  \mathrm{const.} =1$}, {which is shown in Fig.~\ref{fig:rates} (c)}. Here, the growth rate differences increase with increasing $a_0$ and reach $\unit{25}{\%}$ for the longest wavelengths simulated, which is a considerable reduction.

	These changes in the growth rates can be traced back to the collision operators for particle species $q$, summed over all polarization degrees, $\mathcal C_q \equiv \sum_{s} \mathcal C_q^s$. The operators $\mathcal C_q$ govern the quantum transitions for the distribution function of a particle species summed over all polarization states $f_q = \sum_s f_q^s$. We now isolate the polarization dependent terms in the collision operators $\mathcal C_q$ and pinpoint the differences to the \emph{unpolarized collision operators} $\mathcal C_0^\mathrm{unpol}$ used so far for the description of unpolarized QED cascades.
	
	The unpolarized collision operators $\mathcal C_q^\mathrm{unpol}$ contain only the polarization averaged rates of photon emission and pair production,
	\begin{align}
	w_{q} \equiv \frac{1}{2} \sum_{ss'j}  w_{q}^{ss'j}
	\end{align}
	and their momentum-integrated counterparts $W_q = \int_0^k \! \ud p \, w_q (k\to p)$.
	Until now, only the unpolarized collision operators $\mathcal C_q^\mathrm{unpol}$ have been used in simulations of QED cascades {almost exclusively}, see e.g.~Refs.~\cite{Nerush:PhysPlas2011,Elkina:PRSTAB2011,Neitz:PRL2013,Bulanov:PRA2013,Niel:PRE2017}, {with the exception of Ref.~\cite{King:PRA2013} where the effect of photon polarization was included.} {However, they represent only a part of the full collision operators when all the particle polarization is taken into account.}

	The derivation of the full $\mathcal C_q$ from the expressions in Eqs.~\eqref{eq:collision-operator-e}--\eqref{eq:collision-operator-g} is lengthy, but straightforward, and the final result reads
	\begin{align}
	\mathcal C_{-1}(p) &= \mathcal C_{-1}^\mathrm{unpol}(p)  - \Pi_{-1}(p) V_{-1}(p) 
	+  \int \! \ud k \frac{k}{p} \Pi_{0}(k) \, v_{0}(k\to p) \nonumber \\ 
	& \qquad \qquad \qquad \qquad \qquad \qquad 
	+  \int \! \ud k \frac{k}{p} \Pi_{-1}(k) \, v_{-1}(k\to k-p)
	\,, \\
	\mathcal C_{+1}(p) &=
	\mathcal C_{+1}^\mathrm{unpol}(p)  - \Pi_{+1}(p) V_{+1}(p) 
	+ \sum_{q=0,+1} \int \! \ud k \frac{k}{p} \Pi_{q}(k) \, v_{q}(k\to p) \\
		\mathcal C_0(k) &= \mathcal C_0^\mathrm{unpol}(k)  - \Pi_0(k) V_0(k) 
	+ \sum_{q=\pm1 } \int \! \ud p \frac{p}{k} \Pi_q(p) \, v_q(p\to p-k) \,.
	\end{align}
	We see that the difference between the $\mathcal C_s$ and the unpolarized collision operators $\mathcal C_0^\mathrm{unpol}$ contain terms that couple the polarization imbalances of the lepton, $\Pi_{\pm 1} = f^\uparrow_{\pm1} - f^\downarrow_{\pm 1}$, and photon, $\Pi_0 = f^\parallel_{0} - f^\perp_{0}$, distribution functions to the rates' polarization disparities $v_q$. The latter are defined as
	the difference of the scattering rates between the initial particle polarizations, summed over all final state polarization
	\begin{align}
	v_{\pm1}(p'\to p ) &= \frac{1}{2} \sum_{ss'j} s \, w_{\pm1}^{ss'j}(p'\to p) = \frac{1}{2} \sum_{s'j} \left( w_{\pm1}^{\uparrow s'j} - w_{\pm1}^{\downarrow s'j} \right)  \,, \\  
	v_{0}(k\to p ) &= \frac{1}{2} \sum_{ss'j} j \, w_{0}^{ss'j}(k\to p) 
	= \frac{1}{2} \sum_{ss'} \left( w_{0}^{ss'\parallel} - w_{0}^{ss'\perp} \right)\,,
	\end{align}
	with 
	$s= +1 \Leftrightarrow \: \uparrow $, 
	$s=-1\Leftrightarrow \:\downarrow$, 
	$j= +1 \Leftrightarrow \: \parallel$, and 
	$j = -1 \Leftrightarrow \: \perp$.
	Moreover, we defined the corresponding total rate disparities as, e.g. $V_0(k) = \int \! \ud p \, v_0(k \to p)$.

    In our model with polarization taken into account, the rate disparities $v_q$ couple to the particle polarizations $\Pi_q$,
	and this affects even the polarization summed collision operators and therefore causes a change in the growth rates.

    \begin{figure}
    \centering
    \includegraphics[width=0.8\columnwidth]{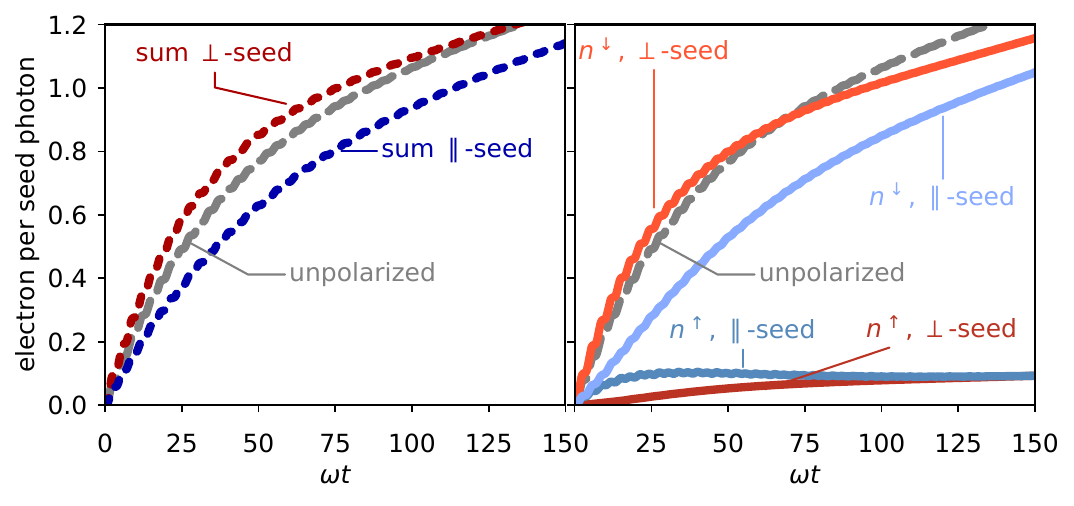}
    \caption{Time evolution of the electron yield of photon seeded cascade for $a_0=400$ and seed photon momentum $k = 10^3 \, mc $,
    with the seed photon initially either $\parallel$- or $\perp$-polarized.}
    \label{fig:polarization_phots}
    \end{figure}

    {Seeding for the cascade is an important consideration \cite{Jirka:PRE2016,Grismayer:PRE2017,Tamburini:2017}.} With the methods developed here, we can also study the evolution of a QED-cascade with an \emph{initially polarized} seed particle distribution
    to make an experimentally verifiable prediction. We choose to examine a polarized photon seed, since polarized photons could be readily generated and their polarization controlled by inverse Compton scattering \cite{Tang:2020}. The basic idea could, however, equally work with an initially polarized electron seed.

    These calculations were performed with $a_0=400$ and a seed photon momentum $k = 10^3 \, mc $, with the seed photon initially either $\parallel$- or $\perp$-polarized. 
    {Fig.~\ref{fig:polarization_phots} (a) shows the time evolution of the particle yields, depending on the polarization of the seed photons. It shows a significantly higher yield of produced pairs for $\perp$-polarized seed photons in the early stage.}

    {Moreover, the electron spin distribution depends strongly on the seed photon polarization, see Fig.~\ref{fig:polarization_phots} (b).} $\perp$-polarized photons produce polarized pairs directly, with a very high abundance of down electrons (up positrons). By contrast, $\parallel$-polarized photons produce equal amounts of up and down electrons and only later are they self-polarized via spin-flip photon emission. For $\perp$-polarized seeds the yield of down electrons alone exceeds the total electron yield predicted in an unpolarized calculation.
    
    These findings suggest an intriguing experimental scheme to investigate the polarization dependence in strong-field QED-cascades with soon-to-be available lasers: Two counter-propagating, circularly-polarized tightly focused 10 PW laser pulses collide and set up a standing wave. A highly polarized gamma-photon beam is injected into the magnetic nodes of the plane wave where the field is a rotating electric field. {Rotating the polarization of the gamma-rays from parallel to perpendicular to the axis of the standing wave will lead to measurable variations in the particle yield, see Fig.~\ref{fig:polarization_phots} (a)}.

	\section{Conclusions}
    \label{sect:conclusions}

    In conclusion, we have shown that QED-cascades will lead to highly polarized particle generation and that the growth rates are reduced by the spin-dynamics, leading to orders of magnitude differences in particle yield compared with calculations with unpolarized rates under certain conditions.
    This raises the prospect of generating polarized lepton beam perhaps useful for laser-wakefield driven particle colliders \cite{Moortgat-Pick:PhysRep2008}, or the production of highly polarized $\gamma$-photons \cite{Li:PRL2020,Tang:2020}. However, addressing the fundamental role of particle polarization in QED cascades, as we have here, could have wide-ranging implications. For example, QED cascades may occur in extreme astrophysical environments such as magnetars and are expected to dominate the behaviour of upcoming high intensity laser interactions with matter as intensities increase and we move into the QED-plasma regime. We have shown that polarization dynamics, which is usually neglected, must be included in the modelling of this state.
    Note that these calculations were performed using an idealized rotating electric field model, which allows to describe the lepton spin in terms of discrete states with respect to a global non-precessing basis.
    Our model does not include a magnetic component or the field gradients expected in near-future high-power laser experiments. In that case, the 3D field configuration requires to generalize the description of the spin degrees of freedom to the full three-dimensional mean spin vector (Stokes vector). It is left for further work to explore the impacts of these effects on the interaction, and how the 3D effects impact the effects predicted in this manuscript.

    This work supported by the US ARO Grant no.~W911NF-16-1-0044, AFOSR Grant No. FA9550-16-1-0121, NSF grant 1804463 and UK Engineering and Physical Sciences Research Council grant EP/M018156/1. D.S. acknowledges valuable discussions with Ben King and Felix Karbstein.

	\appendix
	
	\section{Boltzmann Equation}
	\label{app:A}
    
    {
    In this Appendix we outline the derivation of the Boltzmann kinetic equations for polarized particles, Eq.~\eqref{eq:Boltzmann-normalized}.
    For charged spinless or unpolarized particles the relativistic Vlasov-Boltzmann transport equation is given by
    \footnote{Note that we use un-normalized units in this Appendix.} 
    \begin{align} \label{eq:Boltzmann.unpol}
	\left( \frac{\partial}{\partial t} + \frac{\vec p}{\varepsilon}\cdot \nabla_{\vec x} +  q e  ( \vec E + \frac{\vec p}{\varepsilon} \times \vec B) \cdot \nabla_{\vec p} 
	\right)  f_q (\vec x,\vec p , t) = \mathcal C[f_q,f_{q^\prime},f_{q^{\prime\prime}}\dots]  \,,
	\end{align}
	for the phase space distribution function $f_q$ for species $q$, where $q=\pm1$ represents the charge of the particles, depending on position $\vec x$ and momentum $\vec p$ and dynamically evolving in macroscopic (statistically smoothed) fields $\vec{E}$ and $\vec{B}$, with $\varepsilon=(\vec p^2+m^2)^{1/2}$. The collision operator $\mathcal C[f_q,f_{q^\prime},f_{q^{\prime\prime}}\dots]$ describes brief, short-range interactions between particles of species $q$ and with other species $q^\prime$, $q^{\prime\prime} \dots$ etc. which occur over time- and space-scales negligible compared to those of the macroscopic fields and are therefore treated independently (separation of scales). 
    The kinetic equation for (unpolarized) photons has the same structure, by taking into account that their charge $q=0$ and mass $m=0$.
    
    The collision operators relevant to QED cascades are the strong-field QED processes of nonlinear Compton scattering and nonlinear Breit-Wheeler pair production, which represent the emission of a high-energy photon by a lepton and decay of a photon into a electron-positron pair respectively and are mediated by the classical background electromagnetic fields. The rates for these processes are taken in the locally-constant field approximation (LCFA; see for instance Refs.~\cite{Elkina:PRSTAB2011,Nerush:PhysPlas2011} for explicit expressions). It is assumed that the interactions are collinear, i.e.~the photons are emitted with their momentum parallel to the emitting particle. The LCFA is usually a good approximation for ultrarelativistic particles, but see also the discussion below in \ref{app:rates}.
    We assume that direct high-energy particle-particle interactions, such as M{\o}ller/Bhabha scattering or spin-depolarizing collisions, as well as photon absorption and annihilation processes are negligible due to the relatively dilute high-energy particle density.
    Such processes are usually neglected in QED cascade simulations since they are assumed to become important only at very late times when the produced plasma becomes dense enough \cite{Gonoskov:PRE2015,Blackburn2020,Ilderton:PRD2019b}.

	In order to solve the equations governing the QED cascade dynamics for polarized particles, for the sake of convenience,
	we immediately specify the case of a rotating electric field,
	which is a commonly used model \cite{Bell:PRL2008,Fedotov:PRL2010,Nerush:PhysPlas2011,Elkina:PRSTAB2011,Blinne2014,Mironov2014,Kohlfurst:PRD2019}, and allows for considerable simplifications. In particular, the advantage of this particular choice of system is that the spin states may be defined projected on a non-precessing basis. A path towards a possible generalization to arbitrary field configurations will be discussed below,
	where we will also comment on the limitations of applicability of our model.

	In this work, for electrons and positrons circulating in a rotating electric field, $\vec E = E_0 (\cos \omega t, \sin\omega t , 0)$ and $\vec B=0$, with momentum $\vec p = (p_x,p_y,0)$, there exists a global non-precessing spin basis, parallel to $\vec p \times \vec E$, i.e. the $z$-axis, which is the direction of the magnetic field in the (instantaneous) rest frame of the particles {\cite{Seipt:PRA2018}}, and which we choose as spin quantization axis. A particle initially in a state with its spin (anti)-aligned with that direction  remains in that state during its classical propagation. In the kinetic approach, it is therefore suitable to consider two distributions functions $f_q^s$ for each value of $q$, where $s=\pm1$ distinguishes their spin state as up or down with regard to the nonprecessing axis. Thus, the Vlasov equation for polarized leptons in a rotating electric field in the absence of the quantum interactions reads
	\begin{align}
	\left( \frac{\partial}{\partial t}  +  q e \vec E \cdot\nabla_{\vec p} \right) f_{q}^s = 0\,.
	\end{align}

    For photons, the two independent polarization states are $j=\perp$ (polarization vector perpendicular to the plane of $\vec E$) and $j=\parallel$ (polarization vector lies in the plane in which $\vec E$ rotates). The polarization four-vectors associated to those polarization states
	are $\Lambda^\mu_\parallel = F^{\mu\nu} k_\nu  / | F^{\mu\nu} k_\nu| $ and $\Lambda^\mu_\perp = \tilde F^{\mu\nu} k_\nu  / | \tilde F^{\mu\nu} k_\nu|$, which are by construction transverse to the photon four-momentum $k^\mu$, i.e.~$\Lambda^\mu_j k_\mu = 0$
	\cite{Baier:JETP1975}. $\tilde F^{\mu\nu}$ is the dual electromagnetic field strength tensor.
    These polarization vectors $\Lambda_j$ are eigenvectors of the photon polarization tensor \cite{Karbstein:PRD2013,King:PRA2016}.
    The corresponding photon transport equation is just $\partial f_0^j/\partial t = 0$. The quantum processes can be included in the dynamical evolution of $f_q^s$ by the addition of appropriate collision operators to the right-hand-side.

    The collision operator for polarized particles has the same general structure and approximations as the corresponding unpolarized collision operator \cite{Elkina:PRSTAB2011,Nerush:PhysPlas2011}.
    However, we now use the spin- and polarization resolved rates for nonlinear Compton scattering and nonlinear Breit-Wheeler pair production within the locally-constant field approximation. One could suspect that the collision operator has to take into account {the possibilities} that spin {polarization (Stokes) vector} can be oriented into an arbitrary direction after a quantum transition. This is in fact not the case. It is sufficient to consider {only the component of the leptons' Stokes vector along the non-precessing axis. In this situation the lepton polarization properties can be represented by a mixture of the spin states discussed above,} and we only have to consider transitions between those discrete spin states.
    This can be seen directly by investigating the form of the spin-and polarization dependent rates, which have been recently calculated for arbitrary directions of the polarization directions of all involved particles \cite{Torgrimsson:2020}.
        
    {Let us first assume that, at the moment of photon emission, the electron Stokes vector components perpendicular to the spin quantization axis (SQA) vanish. That means the initial electrons for the scattering are either unpolarized or have a certain degree of polarization along the magnetic field in the particle's rest frame as defined above.} The electron collision operator for Compton scattering must contain the rates summed over the final photon polarization states, {see Eq.~\eqref{eq:collision-operator-e}}. It can be shown by direct calculation that the scattered electrons {Stokes vector in this case is also strictly pointing along the SQA \cite{Seipt:PRA2018,Torgrimsson:2020}. Thus, the leptons tend to polarize along the SQA, and their Stokes vector will not develop other components perpendicular to the SQA. Thus, their distribution can be described by two functions $f^{\pm1}$, with the degree of polarization given by $f^{+1}-f^{-1}$.
	(Because in a rotating electric field the SQA is globally non-precessing the Stokes/polarization vector will remain aligned along that axis.) Moreover, it is known from the literature \cite{Seipt:PRA2018,Torgrimsson:2020}, that the probability of finding an electron in an up or down spin state parallel to the SQA is completely independent of the Stokes vector components perpendicular to the SQA. Thus, the interesting polarization dynamics along the SQA is completely decoupled from the 
	polarization perpendicular to the SQA, and it is by no means necessary to treat the latter explicitly,
	even for arbitrarily polarized initial electrons. Similar calculations can be performed for the contribution of the Compton scattering rates to the photon collision operator, where one has to sum over the final lepton spin states. It reveals leptons polarized along the SQA produce photons in one of the polarization states $\Lambda_j$ discussed above.
	Due to crossing symmetry of the strong-field S-matrix elements, the same arguments must hold for the parts of the collision operator describing nonlinear Breit-Wheeler pair production as well.} We thus can conclude that it is sufficient {and consistent} to consider, also in the quantum collision operators, only lepton the polarization degree along the SQA, and photons being polarized along the directions $\Lambda_j^\mu$.

	In summary, in a rotating electric field the Boltzmann equations for the  six distribution functions $f_q^s$ for polarized particles can be compactly written as
	\begin{align}
	\label{eq:Boltzmann-S}
	\left( \frac{\partial}{\partial t}  +  q e \vec E \cdot\nabla_{\vec p} \right) f_{q}^s = \mathcal C_q^s[\{ f_{q'}^{s'} \}] \,,
	\end{align}
	with $q\in(-1,0,1)$ is the charge denoting different particle species, and $s$ distinguishing the polarization states.
    The collision operators $\mathcal C^s_q$ describe all relevant strong-field QED processes for polarized particles, and possible transitions between polarized particle species: polarized photon emission by polarized leptons and pair production by polarized photons. 
    The explicit form of the collision operators is given in Eqs.~\eqref{eq:collision-operator-e}--\eqref{eq:collision-operator-p}. The {LCFA} transition rates in the collision operator are all functions of their respective quantum parameters
	\begin{align}
	    \chi_q & = \frac{e}{m^{3}}\sqrt{ \varepsilon^2 \vec E^2 - (\vec p\cdot \vec E)^2  } = \frac{e|\vec E|}{m^3} \sqrt{ q^2m^2 + p^2 \sin^2\varphi}\,,
	\end{align}
	where $\varphi = \angle (\vec E,\vec p)$ is the angle between the particle momentum and the instantaneous direction of the electric field vector.
	Going to normalized parameters yields Eq.~\eqref{eq:chi}.
    Polarized leptons are produced by the collision operators with their spins aligned to the global non-precessing SQA axis $\vec e_z$. Thus, the spins of the produced leptons do not precess according to the T-BMT equation \cite{Bargmann:PRL1959} in the rotating electric field configuration.
    Spin-gradient (Stern-Gerlach) forces are not present in the rotating electric field model.

	For general field configurations, there is no global non-precessing spin quantization axis. The model to be used for generalized fields would
	be similar to the quasiclassical approach used for particle accelerators with inhomogeneous magnetic fields: Point-like local quantum transitions causing spin flips (and pair production) with arbitrary direction of the spin, with the distribution precessing classically otherwise \cite{Mane:RPP2005,Derbenev:JETP1973}.
	The classical part of the kinetic equation for relativistic spin-1/2 particles in an extended phase space $f (\vec x,\vec p, \vec s , t)$, depending also on the (continuous) spin variable $\vec s$ can be given as
	\begin{align}
	\label{eq:Boltzmann-S-Ekman}
	\left( \frac{\partial}{\partial t} + \frac{\vec p}{\varepsilon}\cdot \nabla_{\vec x} +  q e  ( \vec E + \frac{\vec p}{\varepsilon} \times \vec B) \cdot \nabla_{\vec p} 
	+ \frac{ q e }{ \varepsilon} 
	\left[ 
	\vec s \times \left( \vec B - \frac{\vec p\times \vec E}{\varepsilon+m} \right)
	\right] \cdot
	\nabla_{\vec s}
	\right) f_q (\vec x,\vec p, \vec s , t) = 0  \,.
	\end{align}
    Here the transport operator in first line is the same as in Eq.~\eqref{eq:Boltzmann.unpol}, the term given in the second line describes the spin precession. In general fields the precession introduces advection in the spin-sector of the extended phase space, i.e.~coupling different $\vec s$.
    Equation \eqref{eq:Boltzmann-S-Ekman} coincides with quasiclassical part of the quantum kinetic equation in Ref.~\cite{Ekman:PRE2017}, where it was derived as the ``spin transform'' of the Wigner function \cite{Vasak:AnnPhys1987,Weickgenannt:PRD2019,Gao:JMPA2020}.
    \footnote{{The $\mathcal O(\hbar)$-terms reported in Ref.~\cite{Ekman:PRE2017} are the quantum corrections to the interaction of the fermions with long-range fields that, e.g., couple field gradients to the spin-dynamics giving rise to Stern-Gerlach type forces, but do not include the quantum transitions due to nonlinear Compton scattering and pair production. By comparing the magnitude of the coefficients of the operators of the quantum corrections it can be argued that these
    extra terms are negligible under the conditions 
    $E m /(E_{S} \varepsilon) \ll 1$, and $\Delta R_F \gg \lambda_C$. The first condition is easily fulfilled since we assume that the field strength $E 
    < E_{S}$, where $E_S$ is the Schwinger field strength, and the particles are ultrarelativistic. The second condition requires that the spatial inhomogeneities $\Delta R_F$ of the electromagnetic field have length scales much larger than the Compton wavelength $\lambda_C$, which is also typically fulfilled for high-power laser plasma interactions. These estimates can also be derived directly from the quantum transport equations in the Wigner formalism, as presented e.g.~in Ref.~\cite{Gao:JMPA2020}.}}
    {As such, the spin precession term in Eq.~\eqref{eq:Boltzmann-S-Ekman} involves a gyromagnetic ratio of 2, consistent with the Dirac equation. It it straightfoward to phenomenologically include also the anomalous magnetic moment $\mu_e$ stemming from loop corrections \cite{Li:PRL2019,Ilderton:PRD2020}. Special algorithms for the accurate simulation of nine-dimensional phase space evolution
    have been developed, see e.g.~\cite{Li_9d_phasespace}.
    Spin-precession can lead to a depolarization of the distributions if individual electrons precess incoherently.
    For instance, in Ref.~\cite{Thomas:2020} a worst case scenario for the depolarization time $T_D$ was estimated as $\omega T_D\sim \pi/[(6\mu_e+4/\gamma) a_0]$, but it is argued that $T_D$ could be much larger for field configurations with certain symmetries.
    In order to estimate the effect of depolarization we have run a numerical simulation for parameters of Fig.~\ref{fig:rates} and $a_0 = 600$ with the lepton distribution forcibly depolarized every timestep, as a representation of the worst case scenario that spin-precession completely depolarizes the lepton distributions. The growth rate reduction calculated with these parameters was 7.6\% instead of 8.3\%, a small effect.}

    The quantum collision operator in the extended phase space in general must also include quantum spin-flip transitions between arbitrary spin-states. {Here one has to describe the dynamics of all components of the lepton Stokes vector during quantum transitions consistently since all Stokes vector components are coupled due to the classical spin precession.}
  	Thus, collision integrals' kernels then must containing the spin-and polarization resolved LCFA rates for the nonlinear Compton and Breit-Wheeler processes with arbitrary direction of the polarization of all involved particles \cite{Torgrimsson:2020,Li2020}. The explicit construction is left for future work.
    }


	\section{Completely Polarized Rates for Quantum Processes}
	\label{app:rates}
	
	Here we provide the full analytic expressions for the fully polarized rates of photon emission and pair production  that enter the collision operator \eqref{eq:collision-operator-e}--\eqref{eq:collision-operator-g}. The rates noted here are the probabilites for the process to happen per unit normalized time $\omega t$.
	
	The fully polarized differential photon emission probability
	for an electron with normalized momentum $p$ to emit a photon with momentum $k$, and go to a momentum state $p'=p-k$ is given by \cite{Seipt-King}
	\begin{multline}
	w_{-1}^{ss'j}(p \to p' ) 
	= - \frac{\alpha {a_S} }{4  p^2}
	\left[ 
	\{ 1+ ss' +jss' (1-g) )\} \: \Ai_1(z)  \vphantom{\frac{1}{1}}\right. 
	\\
	 + 	\left.
        \left\{ y s + u s' + j (u s + ys') \right\} \frac{\Ai(z)}{\sqrt{z}}
      + \left\{  g + ss' + j  \frac{1+gss'}{2} \right\} \frac{2\Ai'(z)}{z}
	\right]\,, 
	\end{multline}
	where {$\alpha$ is the fine structure constant}, $y  = k/p = 1 - p'/p$,
	$u=y/(1-y)$, and $g= 1 + uy/2$.
	$\Ai$ is the Airy function with argument $z= (u/\chi_{-1})^{2/3}$, $\Ai'$ its derivative and $\Ai_1$ the integral $\Ai_1(z) = \int_z^\infty \! \ud x \, \Ai(x)$.
	$j$ is the photon polarization index, $j=+1$ for the $\parallel$ state and $j=-1$ for the $\perp$ state. For the electron spin-states $s=+1$ for up electrons and $s=-1$ for down electrons.
    For instance, $w^{\uparrow\downarrow \perp}_{-1}$ is the probability rate that an up electron emits a photon that is polarized perpendicularly to the electric field in a spin-flip transition going to a spin down state. The rates for positrons emitting photons in Eq.~\eqref{eq:collision-operator-p} can be obtained from the electron rates by inverting all lepton spin variables, $w^{ss'j}_{+1}(p\to p') = w^{\bar s \bar s'j}_{-1}(p\to p')$.
    {The overall prefactor differs from Ref.~\cite{Seipt-King} because here the rate refers to the probability per unit normalized time instead of laser phase, and a different normalization of momenta.}

	The rates for the non-linear Breit-Wheeler pair production by a high-energy photon with momentum $k$ in a polarization state $j$ are given by \cite{Seipt-King}
    \begin{multline}
	w_0^{ss'j}(k\to p) 
	=     \frac{\alpha {a_S}  }{4  k^2}
	\left[
	\{ 1 + ss' + jss' (1 - \tilde g)   \} \: \Ai_1(\tilde z)
	\vphantom{\frac{1}{1}}\right.  
	 \\
    +
	\left. 
	\left\{ \frac{s}{r} - \frac{s'}{1-r} 
	+ j \left( \frac{s'}{r} - \frac{s}{1-r} \right) 
	\right\}
	\frac{\Ai(\tilde z)}{\sqrt{\tilde z}}
	+	
	\left\{ ( \tilde g + ss' ) + j \frac{1+\tilde g ss'}{2} \right\} \frac{2\Ai'(\tilde z)}{\tilde z} 
	\right] \,,
	\end{multline}
	with the Airy function argument $\tilde z = \left(\frac{1}{\chi_0 r(1-r)} \right)^{2/3}$, $\tilde g = 1 - \frac{1}{2r(1-r)}$,
	with $r = p/k$ with $p$ being the generated positron momentum and $p'$ the electron momentum. The approximate momentum conservation now reads $k=p+p'$. Here, $s=\pm1 $ is the positron spin and $s'=\pm1 $ the electron spin.

	These spin and polarization dependent rates are derived from strong-field QED process in a general plane-wave laser pulse in the Furry picture by applying the local constant field approximation (LCFA)~\cite{Seipt-King}. 
	The LCFA generally is considered a good approximation of the full strong-field QED processes if the formation length for the emission of the photon $\lambda/a_0$ is short compared to $\lambda$. This is usually the case for $a_0\gg 1$, but additional constraints, e.g.~$a_0^3/\chi\gg1$ are required \cite{Dinu:PRL2016}. Moreover, it is known that the LCFA can break down for the emission of low-energy photons as for those the coherence length for the formation of the process is $\sim \lambda$ even for $a_0 \gg 1$. The assumption of localized emission is accurate for the emission high-energy photons with energies $k/p \gtrsim \chi/ a_0^3$ \cite{DiPiazza:PRA2018,Ilderton:PRA2019,Blackburn:PRA2020}. Low-energy photons violating that condition are not of high relevance for the formation of cascades since (i) electrons emitting soft photons lose only very little energy
	and thus radiation reaction effects are described with reasonable accuracy within the LCFA even though the photon number spectrum might not be correct at very low $k$; (ii)
	low-energy photons cannot produce pairs for subsequent generations of cascade particles and thus are expected to affect the growth rates only weakly at most. In addition, collinear emission can be assumed for ultrarelativistic particles with $\gamma \gg 1$ (see also \cite{Blackburn:PRA2020} for a discussion non-collinearity effects and its energy dependence). Strictly speaking, the momentum conservation laws found from analysing the strong-field QED S-matrix elements for the scattering in a plane wave background
	with four-wavevector  $\kappa^\mu$ are exact for three \emph{light-front} components of the respective momenta, $\kappa.p' = \kappa.p + \kappa.k$ and $\vec p'^\perp = \vec k^\perp + \vec p^\perp$. In the ultra-relativistic scattering case energy and linear momentum are typically conserved up to terms of the order $1/\gamma \ll1$,
	and the LCFA rates may be used for ultrarelativistic particles interacting with an arbitrary field since in the particle's rest frame the field is boosted to look almost exactly like a crossed field.

	\section{Numerical Methods}
	
	The Boltzmann-type kinetic equations are solved numerically on a rotating radial momentum-space mesh, with typically $300\times 400$ grid points both $p$ and $\varphi$. The time-evolution is calculated using a time-centered operator-splitting method {\cite{Cheng_JCP_1976}}.
	A time-step of $\Delta t = 0.005$ is used for all simulations presented in the main text.  Numerical convergence of the momentum space discretization and the time step was verified.

	To solve the classical advection we use a semi-lagrangian algorithm  {similar to those developed for 1D1P Vlasov systems \cite{Crouseilles_JCP_2010},} which is very efficient because the characteristics are known analytically, {i.e.} Eqs.~\eqref{eq:characteristics-S1} and \eqref{eq:characteristics-S2}, and the distribution functions $f_q^s$ are constant along the characteristics. For each grid point of discretized momentum space ($p_n,\varphi_k$) we employ the characteristics Eqns.~\eqref{eq:characteristics-S1} and \eqref{eq:characteristics-S2} in order to  determine the origin $(p_n',\varphi_k')$ of the parcel at time $t-\Delta t$ arriving at ($p_n,\varphi_k$) at time $t$. The distribution functions are interpolated onto $(p_n',\varphi_k')$ using bi-cubic splines, where the periodicity in $\varphi$ is ensured.

	The action of the collision operator onto the discretized distribution functions can be described by matrix multiplications after discretizing $\vec p$ onto the radial grid ($p_n,\varphi_k$). Those matrices act only on the magnitude $p_n$; the angle $\varphi_k$ appears only parametrically via the quantum parameters $\chi_q$. Thus, independently for each value of $\varphi_k$ we have $\Delta f_{q}^{s}(p_m) 
	= ( \mathcal C_q^s(p_m))_\mathrm{discr} \Delta t=  \sum_{n,q',s'} ( M_{qq'}^{ss'} )_{mn} f_{q'}^{s'}(p_n) \, \Delta t$.
	For instance, the matrix for electron spin-flip from down to up during photon-emission and the up-up non-flip tansition are given by
	\begin{align}
	(M_{-1,-1}^{\uparrow\downarrow})_{mn} & = \sum_j \Delta p \frac{p_n}{p_m} w_{-1}^{\downarrow\uparrow j}(p_n\to p_m) \\
	(M_{-1,-1}^{\uparrow\uparrow})_{mn}  &= \sum_j \Delta p \frac{p_n}{p_m} w_{-1}^{\uparrow\uparrow j}(p_n\to p_m)
	- \delta_{mn} \Delta p \sum_{s',j} \sum_{k\leq n} w^{\uparrow s'j}_{-1}(p_n\to p_k)\,,
	\end{align}
	for $n \geq m$, and zero otherwise.
	For more details on the discretization of the collision operator see for instance Ref.~\cite{Artemenko:PPCF2019}. Our numerical scheme is charge conserving, i.e.~the total charge $Q = \sum_{s,q} q  \int \ud p \ud \varphi \, p \: f^s_{q}  = const.$
	
	As initial distributions for calculations of electron seeded cascades {(with initially unpolarized electrons)} we used the identical distibutions for up and down electrons, $f_{-1}^\uparrow(p,\varphi,t=0)  = f_{-1}^\downarrow (p,\varphi,t=0) = \mathcal N \times e^{-p^2/2w^2}$, where the normalization constant $\mathcal N$ is chosen such that the initial distributions are each normalized to $1/2$. We did run simulations with width $w=100$ and $w= a_0/4$, with almost negligible differeneces in the cascade evolution. For photon seeded cascades the initial distributions are $f_0^j(k,\varphi,t=0) = \mathcal N e^{-(k-k_0)^2/2 w^2} \, e^{ - 4 \varphi^2 }$, normalized to 1, for either $j=\parallel$ or $j=\perp$, and with $k_0=1000$ and $w=100$.


    \bibliography{references}

\end{document}